\documentclass[prl,superscriptaddress,unsortedaddress,twocolumn,%
  showpacs,preprintnumbers,amsmath,amssymb]{revtex4-1}
\usepackage{graphicx}
\usepackage[breaklinks,colorlinks=true]{hyperref}

\newcommand{\LQCD}{\Lambda_{\text{QCD}}}

\newcommand{\GeV}{\;\text{GeV}}

\newcommand{\Tf}{T_{\text{f}}}
\newcommand{\muB}{\mu_{\text{B}}}
\newcommand{\muS}{\mu_{\text{S}}}
\newcommand{\muQ}{\mu_{\text{Q}}}

\begin{document}
\title{Magnetic Shift of the Chemical Freezeout
       and Electric Charge Fluctuations}
\preprint{RIKEN-QHP-222, RIKEN-STAMP-26}
\author{Kenji Fukushima}
\affiliation{Department of Physics, The University of Tokyo, %
             7-3-1 Hongo, Bunkyo-ku, Tokyo 113-0033, Japan}
\author{Yoshimasa Hidaka}
\affiliation{Theoretical Research Division, Nishina Center, RIKEN, %
             Wako 351-0198, Japan}
\begin{abstract}
  We discuss the effect of a strong magnetic field on the chemical
  freezeout points in the ultra-relativistic heavy-ion collision.
  As a result of the inverse magnetic catalysis or the magnetic
  inhibition, the crossover onset to hot and dense matter out of
  quarks and gluons should be shifted to a lower temperature.  To
  quantify this shift we employ the hadron resonance gas model and an
  empirical condition for the chemical freezeout.  We point out that
  the charged particle abundances are significantly affected by the
  magnetic field so that the electric charge fluctuation is largely
  enhanced especially at high baryon density.  The charge conservation
  partially cancels the enhancement but our calculation shows that the
  electric charge fluctuation and the charge chemical potential could
  serve as a magnetometer.  We find that the fluctuation exhibits a
  crossover behavior rapidly increased for $eB\gtrsim (0.4\GeV)^2$,
  while the charge chemical potential has better sensitivity to the
  magnetic field.
\end{abstract}

\maketitle

\paragraph*{Introduction:}
Magnetic fields provide us with a useful probe to reveal non-trivial
topological contents of the ground state of matter or the vacuum in
quantum field theories.  Dynamics of quarks, gluons, and composites is
also significantly affected by an external magnetic field if its
strength is comparable to the typical scale in quantum chromodynamics
(QCD); i.e.\ $\LQCD\sim 0.2\GeV$.  There are many works dedicated to
strong magnetic field effects in the condensed matter and the neutron
star environment~\cite{Lai:2000at,Miransky:2015ava}, in the early
universe~\cite{Subramanian:2009fu}, and recently, more and more
theoretical and experimental studies have been inspired by a
possibility of gigantic magnetic fields created in the
ultra-relativistic heavy-ion collision~\cite{Huang:2015oca}.  There
are transport model simulations~\cite{Skokov:2009qp,*Deng:2012pc} that
have verified an order estimate by simple classical modeling, leading
to a compact formula~\cite{Kharzeev:2009pj,Fukushima:2011jc}:
$eB(t)=eB_0[1+(t/t_0)^2]^{-3/2}$ with
$eB_0\simeq (0.05\text{GeV})^2(1\text{fm}/b)^2 Z\sinh Y$ where the
atomic number is $Z=79$ in the Au-Au collision and the beam rapidity
can be approximated as
$\sinh Y\simeq \sqrt{s_{_{NN}}}/(2m_N)\simeq 107$ at the top energy of
Relativistic Heavy Ion Collider (RHIC) and
$\sinh Y\simeq 2900$ at Large Hadron Collider (LHC).  The impact
parameter $b$ is of order of a few fm and so the largest magnetic
field reaches as large as
$eB_0\sim (\text{a few GeV})^2$ (i.e.\
$B_0\sim 10^{20}\text{gauss}$) or even larger, which must be the
strongest magnetic field existing in the present universe.  The
life-time parameter $t_0$ decreases with increasing energy as
$t_0\sim b/[2\sinh Y]$ but it is important to note that the decay is
not by exponential but by power and so at later time such as even
$t\sim 10^2 t_0$ the suppression factor is $\sim 10^{-6}$.  Besides,
there might be some additional mechanism to sustain the magnetic field
from back-reaction of charged particles~\cite{Tuchin:2013apa}.
Also, it is worth mentioning that the quark matter ground state could
be ferromagnetic~\cite{Tatsumi:1999ab}, which would further enhance
the initially generated magnetic field.

With such a hope of having a substantial residual portion out of the
primordial magnetic field in the heavy-ion collision, lots of recent
developments in the chiral kinetic theory~\cite{Stephanov:2012ki} and
the anomalous hydrodynamics~\cite{Son:2009tf} have been triggered.
The coupling of almost massless (i.e.\ chiral) fermions and the
external magnetic field is an intriguing setup for the investigation
of exotic phenomena caused by quantum triangle anomalies.  The chiral
magnetic effect~\cite{Fukushima:2008xe} (see Ref.~\cite{Liao:2016diz}
for a recent review) has attracted much interest from a broad range of
physics and the impact has spread over condensed matter physics and
astrophysics~\cite{Yamamoto:2015gzz} (and historically speaking, the
chiral magnetic effect was first recognized in the astrophysical
context; see Ref.~\cite{Vilenkin:1980fu,*Metlitski:2005pr}).  The
fluctuation measurement of the electric charge separation is still
ongoing at RHIC and LHC aimed at establishing the chiral magnetic
effect in the heavy-ion collision, while the Weyl/Dirac semimetals
can be cleaner devices for this purpose~\cite{Li:2014bha}.

In parallel to such efforts to study topological effects, our
understanding on bulk properties of QCD matter at strong $B$ has been
advanced tremendously thanks to QCD-like theory calculations and the
lattice-QCD simulations.  It is an established fact that the chiral
condensate increases with larger $B$ and this feature is commonly
referred to as the magnetic
catalysis~\cite{Klimenko:1990rh,*Klimenko:1992ch,*Gusynin:1994re,*Gusynin:1994va,*Gusynin:1994xp,*Gusynin:1995nb,*Shushpanov:1997sf}.
One of the most striking and profound findings from the lattice-QCD
simulation under large $B$ is what is called the inverse magnetic
catalysis~\cite{Bali:2011qj,*Bali:2012zg,*Bali:2014kia,Levkova:2013qda,*Bornyakov:2013eya}
(which was originally used to refer to a response of high-density
matter~\cite{Preis:2010cq,*Preis:2012fh,Fukushima:2012xw}); with
larger $B$ the chiral crossover takes place at smaller $T$ even though
the chiral condensate at $T=0$ becomes larger.  Usually the QCD vacuum
is regarded as a BCS-like state and the critical temperature is
naturally to be proportional to a $T=0$ condensate.  Thus, all
model-based calculations that favor a larger transition temperature at
stronger $B$ have been challenged by this unexpected finding of the
inverse magnetic catalysis.

There are some speculative scenarios to account for the finite-$T$
inverse magnetic catalysis.  The interaction strength in chiral
effective models may have some non-trivial dependence on $B$ so that
this dependence, if the interaction gets weaker with larger $B$, can
overcome the magnetic catalyzing effect (see Ref.~\cite{Gatto:2012sp}
for a review and references therein).  Such behavior of the
interaction weakened by stronger $B$ may well be consistent with the
asymptotic freedom of QCD if the relevant scale is given by
$\sqrt{eB}$~\cite{Braun:2014fua,Mueller:2015fka}.  In other words, the
inverse magnetic catalysis might be a consequence from the confining
sector in which $B$ eases QCD particles of confining forces.  An
alternative scenario called the magnetic inhibition is rather closed
in the chiral sector.  If the magnetic field is large enough to
resolve the internal fermionic contents of neutral mesons, the energy
dispersion of $\pi^0$ is also dimensionally reduced, which would
destruct the chiral order especially at finite
$T$~\cite{Fukushima:2012kc}.  There are also bag-model analyses of
thermodynamic phase transitions with
$B$~\cite{Fraga:2012fs,*Fraga:2012ev}.
\vspace{0.5em}

\paragraph*{Inverse Magnetic Catalysis with the Hadron Resonance Gas Model:}
None of these model scenarios has been fully justified nor falsified
and all of them suffer model-dependent assumptions.  Fortunately,
however, we have another theoretical tool, that is, the hadron
resonance gas (HRG) model, which is free from parameter ambiguities.
At zero magnetic field ($B=0$) and zero baryon chemical potential
($\muB=0$) it has been well tested that the HRG model reproduces the
lattice-QCD data very nicely up to the crossover temperature where the
HRG thermodynamic quantities such as the pressure, the internal energy
density, and the entropy density blow up.  Interestingly, such a
simple picture of the HRG model has been verified also from the
success of thermal model fit of experimental data in the heavy-ion
collision. In this way the chemical freezeout points have been located
on the phase diagram on the $\muB$-$T$ plane (see
Ref.~\cite{Andronic:2009gj} for a summary of thermal model
implications and Ref.~\cite{Alba:2014eba,*Alba:2015iva} for recent
studies on fluctuations to locate the chemical freezeout points).

\begin{figure}
  \includegraphics[width=0.95\columnwidth]{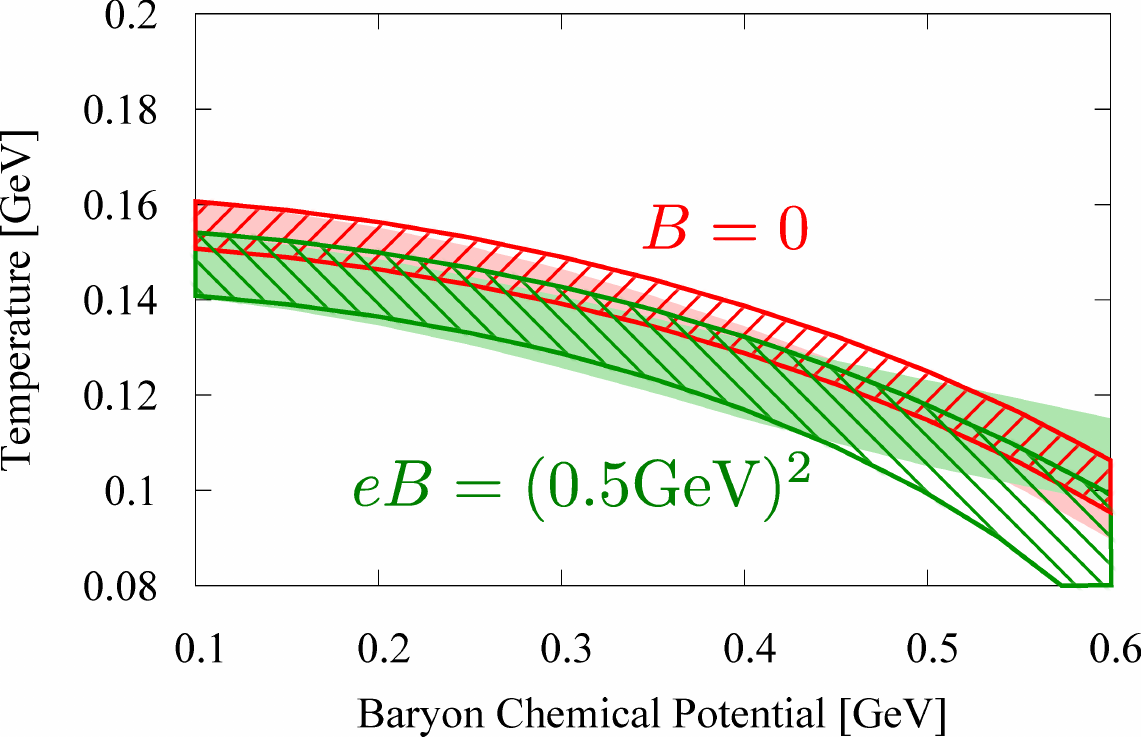}
  \caption{Chemical freezeout bands drawn in the range of
    $E/N=0.9\sim 1.0\GeV$ with and without the magnetic field.  The
    bands with slanting lines represent results with the charge
    conservation taken into account, while the shaded bands represent
    results with $\muQ=\muS=0$ fixed.}
  \label{fig:diagram}
\end{figure}

It has been known that several thermodynamic conditions imposed with
the HRG model can reproduce an experimentally identified curve of the
chemical freezeout~\cite{Cleymans:1998fq,Cleymans:2005xv}.  Among them
a physically reasonable condition is
$E/N=\varepsilon/n \simeq 1\GeV$ where $E$ (and $\varepsilon$) is the 
internal energy (density) and $N$ (and $n$) is the thermal particle
number (density)~\cite{Cleymans:1998fq}.  Here, $N$ counts not only
baryons but also mesons and anti-particles.  Therefore, the chemical
freezeout supposedly occurs when the average energy per one thermal
degrees of freedom (i.e.\ the rest mass plus thermally distributed
energy $\sim m+\frac{3}{2}T$ for non-relativistic heavy particles)
crosses $\sim 1\GeV$.  In Fig.~\ref{fig:diagram} we show bands (with
slanting lines) of the chemical freezeout using the HRG model in the
range of $E/N=0.9\sim 1.0\GeV$ with and without the magnetic field.
In our HRG model treatment we have adopted the particle data group
list of particles contained in the package of
THERMUS-V3.0~\cite{Wheaton:2004qb} (we used only the list and wrote
our own numerical codes).  We should note that we have introduced the
strangeness and the electric charge chemical potentials, $\muS$ and
$\muQ$, to implement the conservation laws of strangeness and electric
charge for the entire system.  More specifically, $\muS$ and $\muQ$
should take finite values to realize $N_S=0$ and $B/(2Q)=1.2683$ where
$B$ and $Q$ represent the baryon number and the electric charge
number, respectively, which is for cold nuclear matter
$(N_{\rm proton}+N_{\rm neutron})/2N_{\rm proton}$ and $1.2683$ is
fixed for heavy nuclei by the $\beta$-equilibrium with the Coulomb
interaction.

\begin{table}
 \begin{tabular}{|c|c|c|c|}
 \hline
 & $a$ [GeV] & $b$ [GeV$^{-1}$] & $c$ [GeV$^{-3}$] \\ \hline
 $B=0$, $E/N=0.9\GeV$ & $0.1519$ & $0.1347$ & $0.05976$ \\
 $B=0$, $E/N=1.0\GeV$ & $0.1618$ & $0.1367$ & $0.04705$ \\
 $eB\!=\!(0.5\text{GeV})^2$, $E/N\!=\!0.9\text{GeV}$
      & $0.1418$ & $0.1253$ & $0.1849$ \\
 $eB\!=\!(0.5\text{GeV})^2$, $E/N\!=\!1.0\text{GeV}$
      & $0.1555$ & $0.1362$ & $0.0565$ \\ \hline
 \end{tabular}
 \caption{Chemical freezeout parameters for $E/N=0.9\GeV$ and $1\GeV$
   with and without the magnetic field.}
 \label{tab:parameter}
\end{table}

The boundaries of the freezeout band (indicated by red lines for $B=0$
and green lines for $B\neq0$ in Fig.~\ref{fig:diagram}) can be
parametrized as a function of $\muB$ in the polynomial form as
$ \Tf(\muB) = a - b \muB^2 - c \muB^4$.
Then, we find that the choice of parameters as listed in
Tab.~\ref{tab:parameter} can give a good fit for the curves in
Fig.~\ref{fig:diagram}.  In fact, for $B=0$, these values determined
from the $E/N$ condition with the HRG model are consistent with the
results, $a=00.166\pm0.002\GeV$, $b=0.139\pm0.016\GeV^{-1}$, and
$c=0.053\pm0.021\GeV^{-3}$, fitted directly with the experimental
data~\cite{Cleymans:2005xv}.

Hereafter we will take the strength of the magnetic field as
$eB=(0.5\GeV)^2$ which may look a bit optimistic estimate but could
possibly be sustained with back-reaction or even strengthened by
ferromagnetism of high density matter.  The condition of $E/N$ with
$eB=(0.5\GeV)^2$ leads to a band shifted down to a lower temperature
as shown by green lines in Fig.~\ref{fig:diagram}.  This clearly means
that the HRG model certainly encompasses the inverse magnetic
catalysis as observed by the chiral condensate in the lattice-QCD
simulation.  We note that Ref.~\cite{Endrodi:2013cs} already addressed
how the HRG model can explain the (inverse) magnetic catalysis, but we
should emphasize that
\textit{it is non-trivial how the HRG model and the inverse
  magnetic catalysis would affect the freezeout curves}.
One would immediately understand this from the shaded bands in
Fig.~\ref{fig:diagram}.  The inverse magnetic catalysis implies that
both $E$ and $N$ rapidly grow up at a lower temperature with stronger
$B$, but it is not obvious which increases faster.  Actually, if the
baryon number density is large, the system is dominated by nucleons,
so that $N$ (or the proton influenced by $B$ directly) increases
faster.  Thus, if the charge conservation is not imposed at high
$\muB$, the chemical freezeout curve as determined by a contour at
constant $E/N$ should be pushed \textit{upward} to a higher
temperature by the $B$ effect, which is indicated by the shaded bands
in Fig.~\ref{fig:diagram}.  Below, we will discuss the effect of the
charge conservation in more details.
\vspace{0.5em}

\begin{figure}
  \includegraphics[width=0.95\columnwidth]{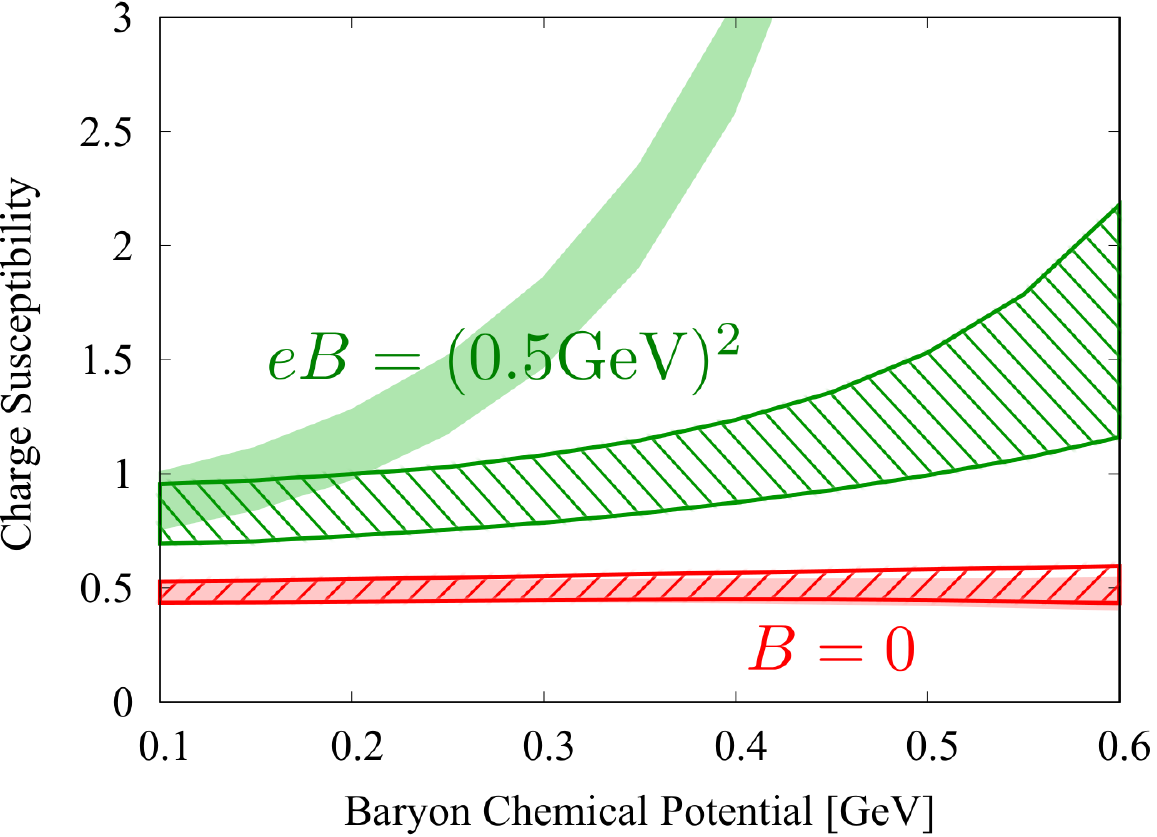}
  \caption{Electric charge susceptibility along the chemical freezeout
    lines, on which the location is indicated as a function of $\muB$,
    with and without $B$.  The bands with slanting lines represent
    results with the charge conservation taken into account, while the
    shaded bands represent results with $\muQ=\muS=0$ fixed.}
  \label{fig:suscept}
\end{figure}

\paragraph*{Conservation Laws and Electric Charge Fluctuation:}
Now that we have identified the chemical freezeout curves on the phase
diagram, we can estimate physical quantities along them predicting
what should be seen in the experiment.  Our central message here is
that the fluctuation of electric charge is quite sensitive to the
presence of $B\neq 0$, where the (dimensionless) electric charge
fluctuation or susceptibility is defined
as $\chi_Q = T^{-2}{\partial^2p}/{\partial \muQ^2}$~\cite{Karsch:2010ck,*Friman:2011pf}.
Figure~\ref{fig:suscept} summarizes our results using the HRG model
with and without $B$ and with and without the charge conservation.
The red band with slanting lines represents $\chi_Q$ at $B=0$ and the
band width corresponds to that in Fig.~\ref{fig:diagram} with two
freezeout conditions $E/N=0.9\GeV$ and $1\GeV$.  The green band with
slanting lines represents $\chi_Q$ at $eB=(0.5\GeV)^2$.  A magnetic
field increases $\chi_Q$ even at $\muB=0$, which is testable in the
lattice QCD simulation (see Ref.~\cite{Buividovich:2009wi} for results
consistent with our present calculation).

\begin{figure}
  \includegraphics[width=\columnwidth]{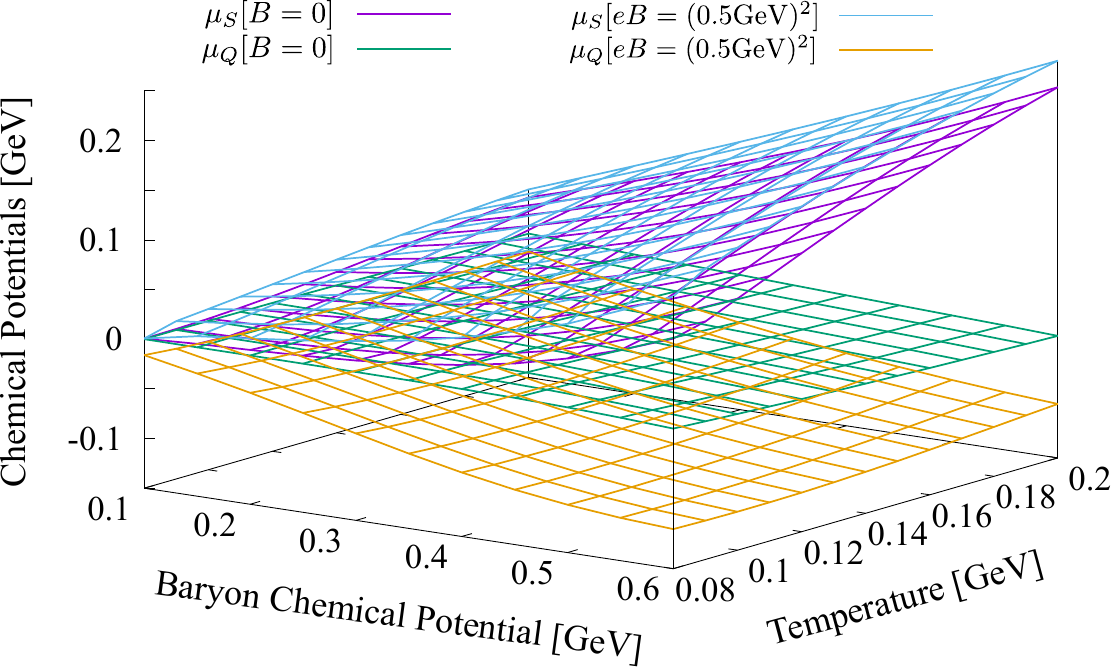}
  \caption{Chemical potentials, $\muS$ and $\muQ$, determined from the
    HRG model and conservation laws, $n_S=0$ and $B/(2Q)=1.2683$, as
    functions of $\muB$ and $T$ with and without $B$.}
  \label{fig:muSQ}
\end{figure}

At larger $\muB$ there are more neutrons and protons in the system.
To keep the ratio, $B/(2Q)=1.2683$, as fixed by the $\beta$-stability
of heavy nuclei, we should introduce a negative isospin chemical
potential or a negative charge chemical potential $\muQ<0$.  As long
as the isospin symmetry approximately holds at $B=0$, the charge
chemical potential $\muQ$ remains of order of $\sim 0.01\GeV$ even at
$\muB\sim 0.6\GeV$ as seen from $\muQ~[B=0]$ in Fig.~\ref{fig:muSQ}.
Therefore, for $B=0$, it should be an acceptable approximation to
neglect the effect of the electric charge conservation at all.
However, the strangeness conservation must be properly implemented
with $\muS$, which is clearly understood from $\muS~[B=0]$ in
Fig.~\ref{fig:muSQ}.  Actually, for $B=0$ and around $T\sim 0.2\GeV$,
we have found $\muS\sim \muB/3$ (= quark chemical potential) as
expected in the deconfined phase.  (We note that the realization of
$\muS\sim\muB/3$ is physically natural, but highly non-trivial in the
HRG model because there is no quark degrees of freedom explicitly
contained in the model.)  Because strange quarks have $S=-1$, the
total chemical potential felt by strange quarks becomes vanishing for
$\muS=\muB/3$, so that the strangeness density is zero then.

Let us consider what happens for a strong-$B$ situation.  Naturally,
the isospin symmetry is explicitly broken, and because the phase-space
density is proportional to the Landau degeneracy factor $eB/(2\pi)$
the proton density is more favored than the neutron density as $B$
increases. Then, there are more protons in the system, causing that
the electric charge fluctuation is also enhanced at stronger $B$, as
is represented by the green shaded band in Fig.~\ref{fig:suscept}.  As
$\muB$ goes up from $0.1\GeV$ to $0.6\GeV$, our calculation shows that
$\chi_Q$ rapidly increases by more than one order of magnitude.
However, such a gigantic enhancement is an artifact from too abundant
protons over neutrons, which violates the electric charge
conservation.  Indeed, for a stronger $B$, as seen from
$\muQ~[eB=(0.5\GeV)^2$] in Fig.~\ref{fig:muSQ}, a larger value of
$\muQ$ of order of $\sim 0.1\GeV$ is necessary to suppress protons
not to violate the charge conservation.  This chemical potential also
suppresses $\chi_Q$ at large $\muB$.  To have an intuitive feeling of
how results change quantitatively, we show all the cases with and
without $B$ and with and without the charge conservation in
Fig.~\ref{fig:suscept} corresponding to the presentation in
Fig.~\ref{fig:diagram}.  We see from Fig.~\ref{fig:suscept} that the
conservation laws have minor effects for $B=0$ (and we can also
understand the strangeness sector hardly modifies $\chi_Q$), but the
proper treatment with the conservation laws is indispensable for
$eB=(0.5\GeV)^2$.  Nevertheless, even after imposing the conservation
laws, we can confirm that $\chi_Q$ at $eB=(0.5\GeV)^2$ is twice
enhanced for small $\muB$ and, for $\muB\sim 0.6\GeV$ it can be four
times as large as that at $B=0$, that implies that $\chi_Q$ could
serve as a probe to detect the presence of a strong magnetic field in
the heavy-ion collision.

\begin{figure}
  \includegraphics[width=0.95\columnwidth]{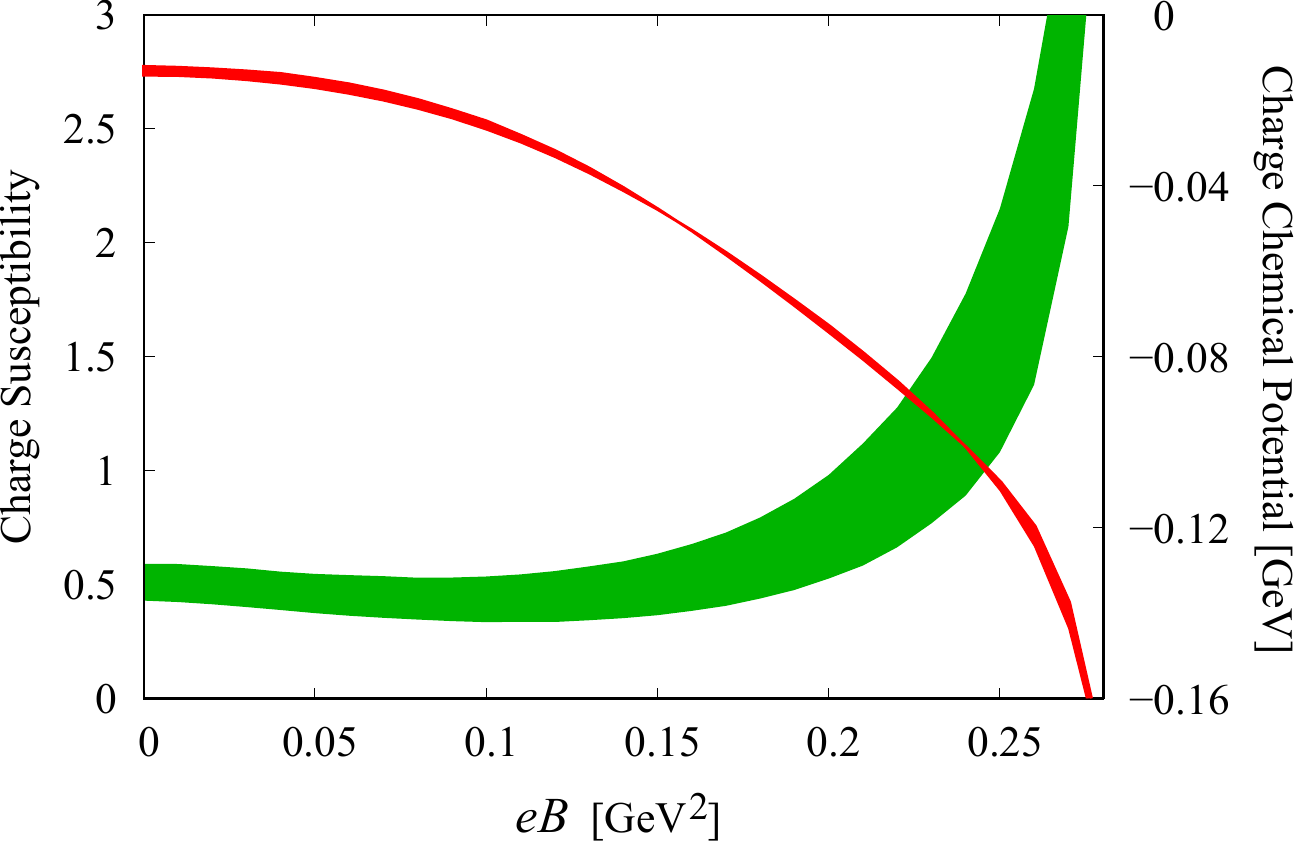}
  \caption{$\chi_Q$ (green line increasing toward the top-right
    corner) and $\muQ$ (red line decreasing toward the bottom-right
    corner) as functions of $eB$ for $\muB=0.6\GeV$.  The
    band corresponds to $E/N=0.9\GeV$ and $1.0\GeV$ and the
    temperature is fixed by respective freezeout conditions.}
  \label{fig:sense}
\end{figure}

We make another plot in Fig.~\ref{fig:sense} to check the sensitivity
of $\chi_Q$ and $\muQ$ as functions of $eB$ for $\muB=0.6\GeV$.  These
are surprisingly non-trivial results;  $\chi_Q$ is almost flat up to
$eB\sim 0.15\GeV^2$ and a quick ``crossover'' of increasing $\chi_Q$
takes place for $eB\gtrsim (0.4\GeV)^2$.  The rapid increasing
behavior in Fig.~\ref{fig:sense} suggests a sort of approximate phase
transition and quark matter at $eB\gtrsim (0.4\GeV)^2$ might be well
identified as a novel state of matter.  It would be an interesting
future problem to explore intrinsic properties of such dense and
magnetized matter.  From the experimental point of view, $\muQ$ should
be a probe that has better sensitivity to the magnetic field.
Practically speaking, moreover, the determination of $\muQ$ would need
less statistics than the fluctuation measurements.
\vspace{0.5em}

\paragraph*{Summary:}
We quantified a shift of the chemical freezeout curves by the magnetic
field effect using the HRG model and the freezeout condition
$E/N\sim 1\GeV$.  We found that the chemical freezeout shift is
downward to a lower temperature in a way consistent with the inverse
magnetic catalysis, though physical contents may be different; we
focused on a combination of $E/N$ and the electric charge chemical
potential $\muQ$ plays a crucial role at large $\muB$ and strong $B$,
not to violate the electric charge conservation.  We proposed an
enhancement of the electric charge fluctuation $\chi_Q$ as a probe to
detect $B$ in the heavy-ion collision.  Alternatively, the proton
number fluctuation could be useful because such an enhancement is
mainly caused by protons whose dispersion should be drastically
modified by the magnetic field.  Also, in principle, $\muQ$ is an
experimentally measurable quantity from the thermal model fit, and
increasing $\muQ$ could serve as a magnetometer.  In particular, we
found that there is rapidly increasing behavior of fluctuation around
$eB\sim (0.4\GeV)^2$ which may signal for a crossover to a yet unknown
new state of matter.

A possible experimental analysis would be as follows.  There are
already nice collections of the thermal model parameters, $T$, $\muB$,
$\muQ$, $\muS$, etc from the beam energy scan program at RHIC and more
data should be expected from the future heavy-ion facilities such as
FAIR, NICA, and JPARC.\ \ These fitted parameters should generally have
centrality dependence, and once this dependence together with the
electric charge fluctuation can be resolved experimentally, our
present calculation suggests that we can extract information on the
magnetic field.  In other words, we emphasize the importance of
\textit{centrality differentiated} measurement of thermal parameters
and fluctuations, and our present study provides future experiments
with a theoretical baseline for the magnetic field detection.
Finally, we would emphasize that, even apart from pragmatic
discussions on experimental opportunities, our present study
sheds a new light of theory on the phase structure and the ground
state properties of highly magnetized QCD matter.
\vspace{0.5em}

\begin{acknowledgments}
  K.~F.\ thanks Vladimir~Skokov and Igor~Shovkovy for discussions.
  This work was partially supported by JSPS KAKENHI Grant
  No.\ 15H03652 (K.~F.\ and Y.~H.),  15K13479 (K.~F.), and 
  16K17716 (Y.~H.)  and partially by the RIKEN interdisciplinary
  Theoretical Science (iTHES) project.
\end{acknowledgments}

\bibliography{mag}
\bibliographystyle{apsrev4-1}

\end{document}